\documentclass[a4paper,twocolumn,oneside,10pt]{mygeneralpapers}
\smartqed
\usepackage{graphicx}
\usepackage{url}
\usepackage{paralist} 
\usepackage{rotating}

\usepackage[normalem]{ulem} 
\usepackage{color}
\usepackage[pdftex]{hyperref}

\journalname{ArXiv CoRR} 
\identifier{} 
\releasedate{14th June 2013} 

\title{Controlled Experimentation in Naturalistic Mobile Settings\thanks{This work was undertaken as part of the RIAM project funded by the EPSRC (EP/E002218/1) and conducted in the Web Ergonomics Lab part of the School of Computer Science at the University of Manchester (UK). About: http://wel.cs.manchester.ac.uk}}

\author{Simon Harper \and Tianyi Chen \and Yeliz Yesilada}

\institute{Simon Harper \at
              University of Manchester, School of Computer Science\\
              Manchester, M13 9PL, UK.\\
              \url{simon.harper@manchester.ac.uk} \\
			  \url{http://simon.harper.name}\\
	          \url{@sharpic}	\\
	          \url{+44(0)161 275 0599}
}

\hypersetup{
	pdfauthor 	= {Simon Harper}, 
	pdftitle 	= {Controlled Experimentation in Naturalistic Mobile Settings},
	pdfkeywords 	= {Small-Device á Mobile Device á Experimental Protocols á Naturalistic Experimentation},
	pdfsubject	= {Performing controlled user experiments on small devices in naturalistic mobile settings has always proved to be a difficult undertaking for many Human Factors researchers. Difficulties exist, not least, because mimicking natural small device usage suffers from a lack of unobtrusive data to guide experimental design, and then validate that the experiment is proceeding naturally.Here we use observational data to derive a set of protocols and a simple checklist of validations which can be built into the design of any controlled experiment focused on the user interface of a small device. These, have been used within a series of experimental designs to measure the utility and application of experimental software. The key-point is the validation checks -- based on the observed behaviour of 400 mobile users -- to ratify that a controlled experiment is being perceived as natural by the user. While the design of the experimental route which the user follows is a major factor in the experimental setup, without check validations based on unobtrusive observed data there can be no certainty that an experiment designed to be natural is actually progressing as the design implies.},
	colorlinks	= true,
	linkcolor	= black,
	filecolor	= black,
	urlcolor		= black,
	citecolor	= black
}

\date{}

\begin{document}
\maketitle
\begin{abstract}~
\begin{description}
\item[Purpose:] Performing controlled user experiments on small devices in naturalistic mobile settings has always proved to be a difficult undertaking for many Human Factors researchers. Difficulties exist, not least, because mimicking natural small device usage suffers from a lack of unobtrusive data to guide experimental design, and then validate that the experiment is proceeding naturally.
\item[Methods:] Here we use observational data to derive a set of protocols and a simple checklist of validations which can be built into the design of any controlled experiment focused on the user interface of a small device. These, have been used within a series of experimental designs to measure the utility and application of experimental software.
\item[Results:] The key-point is the validation checks -- based on the observed behaviour of 400 mobile users -- to ratify that a controlled experiment is being perceived as natural by the user.
\item[Conclusions:] While the design of the experimental route which the user follows is a major factor in the experimental setup, without check validations based on unobtrusive observed data there can be no certainty that an experiment designed to be natural is actually progressing as the design implies.
\end{description}
\end{abstract}
\section{Introduction}\label{intr}
For the Human Factors researcher naturalistic experimentation is key to really understanding users needs~\cite{isomursu2004experience,jambon1993user,kallio2005usability}. However naturalistic methods have one major drawback, that of a lack of control within the experimental procedures. In this case, researchers have turned to controlled experimentation following a reductionist strategy and so, while more accurately quantifiable, these results are often seen as unnatural. Indeed, results gained from experimentation, which does not mimic real life situations, may often produce skewed results~\cite{Gillham:2008tx,De-Vaus:1995ve}. Indeed researchers perceiving this problem, have been working towards methods for adding naturalistic and situational elements to the experimental process\cite{scholtz2001adaptation,waterson2002lab,macefield2007usability}. However, these have mainly been concerned with the specific area of investigation, as opposed to a generic method; and in addition have often been piece-meal approaches (see \S~\ref{back}) not suited to general application. We realise that the use of naturalistic methods in mobile settings is even more critical, but difficult due to the possibility of a high number of confounding variables being introduced to the experimental procedure~\cite{Graziano:2007xr}.

We realised that difficulties exist, not least, because mimicking natural small device usage suffers from a lack of validation checks of naturalistic behaviour to guide experimental design. In this case, our work focuses on the design of these controlled experiments guided by real world unobtrusive observation~\cite{Webb:1966kx} and `ground truthing'\footnote{Ground truth is a term used in cartography, meteorology and a range of other remote sensing techniques in which data are gathered by observation at a distance. Ground truth refers to information that is collected `on location.' The collection of ground truth data enables calibration of data collected unobtrusively, and aids in the interpretation and analysis of what is being observed.} via post observation interviews.

Our observational studies comprised 431 observed participants of which 100 were typing and selecting, and of these, 51 where interviewed. This has enabled us to understand how people interact with their mobile device using either the keypad, pointing device, or a combination thereof, in mobile settings within different environments, locations, and person densities (see \S~\ref{found}). This study allowed us to create seven route scenarios -- chosen based on an aspect of user behaviour changing -- along with four behavioural traits which can be used to monitor the closeness with which experimental behaviour follows naturalistic behaviour as each route scenario is undertaken (see \S~\ref{protocols}). Our route and validation selection lead us to a number of more general points and experiential aspects. These revolved around user task selection, typing and pointing habits, attention switching strategies, mobility condition selection, and the differentiation between public spaces and in-laboratory routes (see \S~\ref{disc}).

Finally, we concluded that researchers following our simple protocols have a much greater chance of uncovering results which more accurately represent small device usage in natural real world settings (see \S~\ref{conc}). We do not claim that our protocols are the finishing point of this type of experimental design, but the starting point, and so our future work not only discusses how we would extend and re-validate our protocols but also how others maybe encouraged to add to them. In the interests of scientific validation full details of the study presented here, including the materials, data, and its analysis can be found as an extensive technical report `How do people use their mobile phones while they are walking? A field study of real-world small device usage', along with the technical report detailing our confirmatory experiments `Investigating Small Device Users' Input Errors under Standing \& Walking Conditions' ~\cite{Chen2009,Harper2011tc} (see \S~\ref{tech}).

\section{Background}\label{back}
Small device user studies are crucial to understand how people interact with small devices and what kind of problems they experience. Without such studies we cannot improve the user experience and advance Human Factors. In the literature, two types of studies have been conducted: laboratory and field studies.

\emph{Laboratory studies} have been widely used mainly because variables can be easily controlled and the protocols are highly reusable. Therefore, they are typically used for hypothesis testing and product evaluation. However, due to the controlled environment, laboratory studies are often criticised for limited realism and unknown generalizability~\cite{kjeldskov2003review}. Laboratory studies with small devices are often conducted in a controlled environment, such as in an office~\cite{Lin2007} or in a hall way~\cite{Bohnenberger2002}. In this case researchers use pre-defined track to control environmental variables. Other researchers minimise the distractions from environment by using treadmills to simulate motion, in which case a large degree of experimental control is achieved at the price of limited relation to the real world~\cite{Barnard2005,Mizobuchi2005,Mustonen2004}.  

On the other hand, \emph{field studies} are normally conducted under ``real-world conditions'' as opposed to in a laboratory setting. Ethnographic field studies are characterised by researchers immersing themselves into the environment they study, gathering ethnographic data through observations and interviews~\cite{Webb2000}. Field experiments apply such ethnographic data to controlled experiments where the influence of environmental variables on dependent variables are observed in a natural setting~\cite{Brewster2002,goodman2004using,Kane2008a}. In comparison with laboratory studies, the major advantage of field experiment is increased realism and generalizability. Field studies have been widely used to investigate the use of small devices~\cite{Kristoffersen1999,Petrie1998,Weilenmann2001}. Kristoffersen~\cite{Kristoffersen1999} conducted two field studies with telecommunication service engineers and maritime consulting staff who were heavily involved in field work and used small devices for receiving orders and communicating with colleagues in the field. Their results illustrated the primary problem field workers faced when using small devices was that the interaction required too much visual attention and it required two hands for input. Pascoe~\cite{Pascoe:2000:UWM:355324.355329} analysed the fieldwork of a group of ecologists observing giraffe behaviour in Kenya. Weilenmann~\cite{Weilenmann2001} conducted a field study with 11 ski instructors during a one-week ski trip. Both studies generated similar results: they found that fieldworkers used a small device in very dynamic context (e.g., while standing, walking, crawling or skiing), with limited attention on the device. They also needed high-speed interaction where the device needed to be able to enter high volumes of data quickly and accurately. In addition, location awareness is also an important feature of the small devices used in outdoor environments. While field studies can generate rich amount of data in relatively short time, the major disadvantage of this method is the unknown bias and uncertainty of the representativeness of the data. It is possible that behaviours of the participants of the field study are specific to certain population and thus hard to generalize~\cite{kjeldskov2004worth,kjeldskov2007studying}.

As can be seen from the discussion above, both laboratory and field studies have pros and cons\cite{bartek2003experience,Bennett:2006:RTH:1125451.1125761,21863,betiol2005usability,brush2004comparison}. Some existing studies tried to combine the advantages of both kinds of studies, however, they do not provide, as we do here, generic protocols that can be used by other researchers. For instance, in Brewster 2002~\cite{Brewster2002}, participants walked up and down a 10 meter long straight pathway in a university campus while conducting tasks with a given PDA device. Once a participant reached the end of the pathway, the participant turned and walked back. They continued such loops until the experiment time ran out.  The task participants conducted was entering a series of five digit codes using a on-screen numeric keyboard. There was no control on a participant' pace and the path was quiet with relatively little distraction to the participant. In Kane 2008~\cite{Kane2008a}, a more complicated route was used. Instead of walking up and down on a short straight pathway, participants were asked to walk on a longer curved path in an open plaza on a university campus. In order to maintain a consistent walking speed among participants, all participants followed an experimenter while walking on the path. Tasks included making selections on a MP3 player and playing music.  Although the setting is more naturalistic, participants still repeated the same path several times. In addition, the difference between following an experimenter and walking by oneself on a participant's performance still needs further investigation. Similarly, Oulasvirta~\cite{Oulasvirta2005} looked at how context affects small device users' attention. In their study, a participant was equipped with three small cameras: two mounted on the mobile phone and one attached to the participant's coat. These cameras were used to record the device screen, the surrounding environment, and the participant's face. In addition, an experimenter carried a fourth camera to record the whole scene. Results of the study showed that when walking in public areas, small device users had much rapid attention switches, and compared with that in a laboratory, the continuous span of attention to the small device was much shorter in public areas. Oulasvirta's~\cite{Oulasvirta2005} study was conducted in public areas without artificial setting. However the participants had to carry additional equipments with them, which would affect their performance.
\section{Controlled Experimentation in Naturalistic Mobile Settings}\label{protocols}
We conducted a field study that investigated the behaviour of small device users in naturalistic settings. We wanted to understand questions such as: Do small device users walk and use the device at the same time? Do they look around while typing or just focus on the device screen with little attention to the surrounding environment? Do they correct their typing errors? Do they use abbreviations? What is the keyboard they prefer? The field study consisted of two stages: a series of unobtrusive remote observations and interviews. In order not to disturb the users and thus alter their behaviours, the experimenter played a passive and non-intrusive role during the observations. This study was performed in five different locations in Manchester, United Kingdom, including a shopping centre, a train station, a bus stop, a business and a market street. In the first stage, 431 small device users in total were observed, 100 of whom were typing on their devices, and the other 331 were making phone calls. In the second stage, a total of 51 small device users were interviewed. The interview study was designed in such a way that observational results were further investigated. On average an interview was lasted between 5 to 10 minutes. The major findings of both studies are detailed in our technical report~\cite{Chen2009} and journal paper~\cite{Harper2011tc}. 

\subsection{Step-by-Step Design}

Based on our field study results, we provide a protocol on designing controlled experimentation in naturalistic mobile settings. In general, the protocol is formed based on the following principles:
\begin{inparaenum}
\item A study should include different topographical conditions. In real world, small device users are likely to use the devices while traversing through areas with different conditions, such as on a straight road, on stairs and at corridors. Comparing with using simple path or treadmills, introducing various topographical conditions will increase realism, thus generate more representative data.  
\item A study should include different mobility conditions. Again, in real world, people use small devices while they are sitting in a park, standing at a bus stop, walking alone or walking with friends. These mobility differences should be coded in to the study so that the performance of small device users under different mobility conditions can be compared. 
\item An experimenter should take note on each participant. Results of our field study show that there are general patterns in terms of typing habits and attention switches of small device users. For example, small device users usually type with just one hand and they normally have rapid attention switches when using the device while walking. Such information can be used as indicators of realism of a controlled experimentation. By collecting and analysing these data, the researcher can ratify, or validate, that the experiment is proceeding naturalistically.  
\end{inparaenum}
With these principles in mind it is possible to derive a set of general steps to follow when planning a new experiment:

\begin{description}
\item[Identify Independent Variables:] Mobility conditions, topographical conditions, and other factors such as congestion level, lighting condition and noise level are independent variables that will affect participants' performance. These need to be clearly identified before the experiment;
\item[Define Values For Each Variable:] If researchers choose mobility condition and topographical condition as two independent variables, they then need to consider possible values for each variable. If the mobility condition has two values (walking alone and walking while accompanied), and topographical condition has three values (open space, straight corridor, stairs), the researchers will have 6 (2$\times$3) combinations;
\item[Choose a Route:] Now the researchers need to choose a route that accommodates the possible combinations of the independent variables. The route should be long enough so that same route will not be repeated during the experiment. In addition, open space, corridors and stairs (see later) need to be included in the route so that the topographical variable can be implemented;
\item[Split The Route Into Segments:] After choosing a path, the researchers now divide the path into six segments, each corresponding to one variable combination. The start and end point of each segment should be clearly identified just to keep consistency among participants;
\item[Choose Tasks:] Task selection is highly dependent on the aim of the experiment. To be realistic, tasks conducted on small devices cannot be too long. People will generally perform better with materials and tasks that they are familiar with. Using day-to-day tasks will recall their daily experiences in using the devices, this generates more realistic results as opposed to using tasks that are less likely to undertaken in real life;
\item[Assign Tasks To Segments:] Now researchers assign one interaction task to each segment. To minimise learning effect, materials used for each task should not be the same. However, the difficulty of the task should be consistent. This can be achieved by maintaining a same ratio of characters to numbers or characters to punctuation marks through each task;
\item[Set Up User Interface:] Now that tasks and route are defined, researchers need to set up interface components for participants use in the experiment. A logging method or tool needs to be in-place so that user interactions can be recorded in real time, the logging should be conducted unobtrusively; and
\item[Assign An Observer:] One principle of our protocol is that researchers keep track of certain indicators of realism. Here the researchers need to assign an observer whose job is to remotely observe a participant during the experiment. The observer take notes based on the validation criteria; as defined below. The observer can keep a short distance behind or ahead the participant, so that he/she will not disturb the participant but sill has a clear view.
\end{description}

\subsection{Components of the Environment}\label{found}
Results of the current field study (see \S~\ref{tech}) suggest that small devices are not only used in static position, but also used while the users are walking. Further, the use of small devices in walking condition is not only limited to phone calls, but also includes other attention demanding task like typing and reading. Therefore, in a user evaluation, the use of a small device should be tested under different mobility conditions, certainly under walking condition. In Lin et al.'s study, small device users' pointing performance was examined under three different conditions: sitting, walking on a treadmill, and walking through a defined route in a laboratory room~\cite{Lin2007}. Further, Kjeldskov and Stage's study also involved six mobility conditions: sitting, walking on a treadmill with constant speed, walking on a treadmill with varying speed, walking on a court at constant speed, walking on a court at varying speed, and walking in a pedestrian street~\cite{kjeldskov2003review}. These settings, although cover most of the mobility conditions, are still unnatural as participants normally conduct different trials, each of which corresponds to a specific mobility condition. We suggest that different mobility conditions can be all coded in a single trial. For example, participants can go through a route which mixes different topographical conditions. They can also stop at certain points to simulate the sitting or standing condition.

Our work has allowed us to identify a set of generic protocols that have much greater chance of uncovering results which more accurately representing small device usage in natural real world settings; the work is based upon two foundations. Firstly, the environment to be traversed is sectioned based on the behavioural differences we have identified which are associated with that environment; for instance traversing stairs is associated with different user interaction behaviours than, say, traversing a corridor. Secondly, each section of the traversed environment has an associated set of behavioural traits which are used as validation that the experiment is proceeding in a naturalistic fashion. The following subsections detail each environment and validation checks that accompany it, along with further details on why it was included and an explanation of its experimental underpinnings. 

\subsubsection{Stationary} 
\emph{A stationary component, either standing or sitting, should be included in any experiment with mobile users}. This stationary aspect is required as mobile users often take the opportunity to stop and focus on aspects of the interaction which are either cognitively complicated or when speed is required.

\paragraph{Validation -- Holding the Device} The observational results showed that majority of the mobile users, observed when stationary, held the device with both hands. This result was also confirmed in the interview study; the majority of the people interviewed they indicated that they typically hold their mobile phones with both hands when stationary and when large amounts of input needs to occur.
\paragraph{Validation -- Entering Data} Our results showed that users hold the device in their palm and pressed the keys with both thumbs. It was rarely observed that people used other fingers for typing. This was also confirmed with the interview; the majority of the people interviewed, they said they typically use their thumbs for pressing the keys; and
\paragraph{Validation -- Attention Switches} Almost all small device users observed had very slow attention switches while stationary. They typically switched their attention  from the device screen to the environment very occasionally. It was also observed that attention switches while typing were less when people were standing or sitting still compared to walking condition. Our interview study also confirmed these findings. Even though when explicitly asked some people said they did not exhibit attention switches, only focusing on their device, the majority of the people interviewed said they do remember switch attention.
\paragraph{Validation -- Interaction Rate} Interactivity is fastest of all when the user is stationary. There are few attention switches, input and pointer control occurs with both thumbs. This means that interaction is fast relative to the other route segments, on a per individual basis.
\subsubsection{Open Spaces (Empty)}
\emph{Participants should traverse an open space which is either empty or lightly populated}. We noticed differences in usage and behaviour when mobile users traversed open areas free of people or obstacles to avoid. In this case, experiments should include these kinds of areas to give more accurate results.

\paragraph{Validation -- Holding the Device} The observational results showed that majority of the mobile users observed, the majority used the device with just one hand. When the users were on the move, they typically used their other hand for carrying bags, holding books or some other goods. This result was also confirmed in the interview study. A majority of the people interviewed indicated that they typically use their mobile phones with one hand.
\paragraph{Validation -- Entering Data}  Our results showed that of those who typed with one hand hold the device in their palm and press the keys with one thumb. It was rarely observed that people used other fingers for typing. This was also confirmed with the interview; The majority of the people interviewed said that they typically use their thumb for pressing the keys.
\paragraph{Validation -- Attention Switches} Almost all small device users observed had attention switches while walking. They typically switched their attention from the device screen to the path that they walked on. When they typed while walking, they normally checked the path ahead, focused on typing, and checked the path again after a few seconds. Our interview study also confirmed these findings.
\paragraph{Validation -- Interaction Rate} Again, interaction rate is fast, because there are less attention switches but the user is still using only one hand, but slower than `Stationary'.
\subsubsection{Open Spaces (Routes)}
\emph{Participants should traverse and open space populated with other people moving in distinct flows}. Think of an open space in which pedestrians naturally group into a lane moving in one direction, while another group into a lane moving in a different direction. In this case the flow resembles a `Follow-My-Leader' traversal and so has different validation behaviours.

\paragraph{Validation -- Holding the Device} As with `Open Spaces (Empty)', the observational results showed that a majority of the mobile users held the device with just one hand. However, in this case, when the users were on the move, they typically used their other hand for security incase the slipped of fell. Again, this result was also confirmed in the interview study.
\paragraph{Validation -- Entering Data} As with `Open Spaces (Empty)', the thumb of the hand holding the device is used.
\paragraph{Validation -- Attention Switches} Almost all small device users observed had rapid attention switches while walking; one every two seconds. They typically switched their attention from the device screen to the path that they walked on. It was also less when people were accompanied with somebody else compared to when they walked alone; in this case being part of the flow elicited behaviour similar to be accompanied but without a trusted accompanying party.
\paragraph{Validation -- Interaction Rate} Interaction rate slows down to similar rates as those associated with `Follow-My-Leader / Walking With a Companion'.
\subsubsection{Open Spaces (Junctions)}
\emph{The route should be through an open space which forms a convergence of routes such that some pedestrians will be stationary}. This kind of environment alters the behaviour of users who start to attention switch more often, and who's interaction rate slows down.

\paragraph{Validation -- Holding the Device}  As with `Open Spaces (Routes)', the observational results showed that a majority of the mobile users held the device with just one hand. As with `(Routes)', they typically used their other hand for security but in this case it was to fend off stationary people.
\paragraph{Validation -- Entering Data} As with `Open Spaces (Routes)', the thumb of the hand holding the device is used alone.
\paragraph{Validation -- Attention Switches} Rapid attention switches increase above one every two seconds. This is because they are really performing collision detection with stationary obstacles and pedestrians.
\paragraph{Validation -- Interaction Rate} Interaction rate is very low, or in some cases stops completely until the users has traversed this section; and the perceived danger of collision is removed.
\subsubsection{Traversing Stairs}
\emph{The route should include the ascent or descent of stairs; although both ascent and descent are not required, just one is sufficient}. Stairs have similar properties as `Open Spaces (Junctions)' because there is a danger associated with both; collision detection occurs in `(Junctions)', while fall prevention occurs in stair traversal.

\paragraph{Validation -- Holding the Device} Again, the observational results showed that a majority of the mobile users held the device with just one hand. As with `(Junctions)', they typically used their other hand for security to guard against falls.
\paragraph{Validation -- Entering Data} Again, the thumb of the hand holding the device is used alone.
\paragraph{Validation -- Attention Switches} Similar attention switches occur to `(Junctions)' because again users are concentrating on safety.
\paragraph{Validation -- Interaction Rate} Interaction rate decreases and in some cases slows to a stop, especially when attention is focused on not tripping.
\subsubsection{Corridors, Curb or Edge Following}
\emph{The route should encompass both corridors and curbs so that the participant has the opportunity to `edge follow'}. Here we see attention switches slowing down, and interactivity measures increasing because users can follow an edge by using their peripheral vision to understand proximity as they are already looking down. They can also tell if a pedestrian is very close by observing their feet, and in the case of a corridor some people relay on others moving out of their way if the mobile users is very close to the corridors wall.

\paragraph{Validation -- Holding the Device} Again, the device is held with just one hand.
\paragraph{Validation -- Entering Data} Data is entered using the thumb of the hand holding the device.
\paragraph{Validation -- Attention Switches} Attention switches decrease to one every four to five seconds and in some cases much longer. For example, we observed a lady (one of many users) using a PDA when walking on the pavement very close to the edge of a road. While using the device and walking, this lady had no attention switches in about 20 seconds, which is much less than the average number of attention switches observed with other small device users. This is possibly because the lady knew that no one was going to collide with her from the opposite direction (because there was not enough space between the curb and herself for anyone to go through), and she could just follow the road by scanning the curb with her foresight while looking at the screen of her PDA. Therefore, the user did not need to deliberately shift her attention for path finding. This suggests that when following an edge, small device users will have less attention switches than walking in open area.
\paragraph{Validation -- Interaction Rate} The rate of interaction shows a marked increase to near `Stationary' levels. In a controlled user evaluation, walls on both sides of a corridor, edge of the tape on a treadmill, and the marks on a clearly marked route may have the same effect to a user as the curb of the road did to our archetypal lady, which serves as a way edge~\cite{Harper2000uq,Weilenmann2001}. In such settings, users know that they will not be disturbed and by scanning the way edge, a participant may save efforts for path finding and focus more on the device screen. 

\subsubsection{Follow-My-Leader / Walking With a Companion}
\emph{The participant should perform interactions while either following a person along a section of the route, or walking with a companion}. 

\paragraph{Validation -- Holding the Device} The device is held with just one hand.
\paragraph{Validation -- Entering Data} Data is entered using the thumb of the hand holding the device.
\paragraph{Validation -- Attention Switches} Switches are less when people were accompanied with somebody else compared to when they walked alone. Our interview study also confirmed these findings. Even though when explicitly asked some people said they did not remember attention switches while following or being accompanied -- only focusing on their device -- a majority, said they often did have attention switches.
\paragraph{Validation -- Interaction Rate} Interaction rate increases and is normally slightly higher than `Corridors, Curb or Edge Following'.

\subsubsection{In Summary\dots}
\begin{table*}[t]
\begin{center}
\begin{tabular}{|l|l|l|l|l|}
\hline
\multicolumn{5}{|c|}{{\bf Route Environments}}\\\hline
~&{\bf Stationary}&{\bf Stairs}&{\bf Edges}&{\bf Accompanied}\\\hline\hline
{\bf Holding}&Both Hands&1 Hand&1 Hand&1 Hand\\\hline
{\bf Using}&Both Thumbs&1 Thumb &1 Thumb &1 Thumb\\\hline
{\bf Switches}&1 / 10 seconds&$>$ 1 / 2 seconds&1 /  5 seconds&$>$ 1 /  5 seconds\\\hline
{\bf Rate}&High&Low&Medium&Medium-High\\\hline\hline
\multicolumn{5}{|c|}{{\bf Open Spaces}}\\\hline\hline
~&{\bf Empty}&{\bf Route}&{\bf Junction}&~\\\hline
{\bf Holding}&1 Hand&1 Hand&1 Hand&~\\\hline
{\bf Using}&1 Thumb &1 Thumb &1 Thumb &~\\\hline
{\bf Switches}&1 / 4 seconds&1 / 2 seconds&$>$ 1 / 2 seconds&~\\\hline
{\bf Rate}&Medium&Medium-Low&Low&~\\\hline
\end{tabular}
\end{center}
\caption{Summary of Route Types with Validation Checks (How are devices held and operated, how many attention switches to check the environment are there, and how fast are user interactions progressing?)}
\label{route-summary}
\end{table*}%
Based on these principles, we suggest that small device user evaluation should distinguish the setting of walking along a clearly marked route and the setting of walking in public space where no specific route is acquired. Using a clearly marked route has the benefit that all participants follow the exactly same route and thus the effect of route variance is fine controlled. However, the trade-off is that the disturbance a user receives is low in these settings and thus the performance of that user may be overestimated. On the other hand, using public space means the setting is more naturalistic and that a user may have more attention switches. In these situations, the observer can note user behaviour and compare it with our validation checks (see Table~\ref{route-summary}) to gauge how `naturally' the experiment is proceeding. Further, our interview study also revealed more findings that could not be obtained in the observational study. According to the interview results, small device users normally typed with the predictive text function turned on; they also used abbreviations in text messages; and correct the typing errors they made, preferring to use a physical keyboard over a soft-keyboard; including these aspects within the experiment will also increase the feeling of naturalness without compromising the controlled nature of the dependent variables.

\section{Confirmatory Studies}\label{prot}
We have previously discussed (see \S~\ref{found}) the methods by which we have derived our experimental protocols and the guidelines that go with them. In effect we can see this as our formative experimentation, which suggests that there is a summative, confirmatory, aspect to this work. In this case we have confirmed that our protocol is fit for purpose by using it to conduct two further experimental trials. These trials were conducted in an iterative fashion such that any additional information or knowledge missed in the first summative work was rapidly fed back into the experimental protocol so that the next naturalistic experimentation, and participant, could benefit from these new lessons.

Here, we worked with two separate group of participants (20 and 15 each) and created an experimental design based upon the protocols derived. This experimental design underwent an ethical review (see \S~\ref{disc}), including a risk assessment of the route, with a number of external interdisciplinary researchers taken from different schools within the University of Manchester. This review found that the protocols were satisfactory both as research methodology and in the ethical sense, and that the only aspect which would need to be reanalysed would be to risk assessment based on changes within the gross route as opposed to a change in order of the different sections of the route. 

Nine tasks summarised in Table~\ref{tab:task_route} are planned on this route. The route was designed around the outside and inside of a building of the University of Manchester. A participant starts at the loading bay and conducts the first task. Then the participant walks on the pavement around the building to the automatic door on the first floor. In the mean time, the participant conducts the second task. Then the participant enters the building and stands by the message board and conducts the third task. After that, the participant walks upstairs and goes to a pre-identified room while performing the fourth task. When arriving this room the participant stands by the door and conducts a fifth task. Upon finishing, the participant walks along the straight corridor and finds an experimenter at the end of the corridor, during which the participant finishes the sixth task. The experimenter then leads the participant back to the automatic door where the participant enters the building. While being lead, the participant finishes the seventh task. Then the experimenter leads the participant to the sign outside of the building, while the participant conducts the eighth task. After that, the experimenter leads the participant back to the loading bay, and the participant conducts the last task.

\begin{table*}[!t]
\begin{center}
\begin{small}
\begin{tabular}{|p{1.5cm}|p{2cm}|p{2cm}|p{2cm}|p{6cm}|}
\hline
\textbf{Task}	&\textbf{Mobility}	&\textbf{Task Type}	&\textbf{Sub-route}	&\textbf{Checkpoint} \\\hline
T1	&standing	&pointing 	&1	&loading bay \\\hline
T2 &walking alone	&typing	&2	&loading bay \\\hline
T3 &standing &editing	&3	&automatic door on the first floor of the building \\\hline
T4 &walking alone &pointing &4 &message board \\\hline
T5 &standing 	&typing 	&5	&door of pre-identified room\\\hline
T6 &walking alone	&editing 	&6	&door of pre-identified room \\\hline
T7 &guided walk	&pointing &7 &meet the experimenter \\\hline
T8 &guided walk &typing 	&8 &automatic door on the first floor of the building \\\hline
T9 &guided walk &editing	&9 &sign of building \\\hline
\end{tabular}
\end{small}
\caption{Task conditions and corresponding sub-routes and checkpoints.\newline For sub-route details see Table~\ref{tab:sub_route}}
\label{tab:task_route}
\end{center}
\end{table*}

\begin{table*}[!t]
\begin{center}
\begin{small}
\begin{tabular}{|p{2cm}|p{12.76cm}|}
\hline
\textbf{Sub-route} &\textbf{Details}\\\hline
1	&Standing at the loading bay area behind the building. \\\hline
2 &Walking from the loading bay area to the automatic door on the first floor of the building, via Oxford Road. \\\hline
3 &Entering building, standing by the message board near the automatic door. \\\hline
4 &Going upstairs to 2nd floor of the building and walking to the pre-identified room. \\\hline
5 &Standing by the door of the pre-identified room. \\\hline
6 &Walking along the corridor, and finding an experimenter in black jacket.  \\\hline
7 &Following the experimenter to the automatic door of building \\\hline
8 &Following the experimenter to the sign outside of building. \\\hline 
9 &Following the experimenter to the loading bay where the study started. \\\hline
\end{tabular}
\end{small}
\caption{Sub-routes in detail}
\label{tab:sub_route}
\end{center}
\end{table*}

We then compared our original experimental designs based on the related work that was available, generated before the observational experimental work and the definition of our protocols, to our final experimental design. In this case we found a number of significant differences, all centred around the naturalistic aspects of the experimental work such as attention switching, mobile usage, topology and environmental conditions. Indeed, these changes led us to believe that there was space within the experimental data in which to place our protocols and guidance. In this way, we can confirm that the experimental design, derived by the application of our protocols, gave benefit in terms of designing controlled experiments in naturalistic settings, which would not have otherwise been achievable.
 
\section{Discussion}\label{disc}
There are a number of other factors which warrant a more general discussion with regard to implementation of experimental design and naturalistic settings. One major area for discussion is that of the ethical framework in which the exponents occur. We see the ethical process is a critical component of good experimental design because it encourages us to focus on the methodology and the analysis techniques to be used within the experimental design. More properly, to make sure that the study-designers possess a good understanding of what experimental procedures will be carried out, how they will be analysed, and how these two aspects may impact the participants, and the resultant science. In reality then, our protocols are also focussed on ensuring due diligence, a form of standard of care relating to the degree of prudence and caution required in experimentation. To breach this standard may mean that the resultant data does not enable us to understand mobile usage in naturalistic settings.

Another key element of good experimentation in naturalistic settings, and in this case we mean for the experimental participants to be mobile, is the risk assessment. This assessment is the determination of risk related to a concrete situation or a recognised hazard. There are a number of factors which affect the possible safety of the experimentation and therefore involve some possibility of risk to the participant. The three main steps involved in the estimation of the risk are: (1) Hazard Identification, determine the nature of the potential adverse consequences; (2) Response Analysis, determine the relationship between the probability or the incident and the effect, taking into account differences between individuals or other factors which may mean that the hazard may be higher for particular groups; and finally, (3) Exposure Quantification, determine the size of the hazard that participants may be exposed to. As the ubiquitous interface becomes more common and the desire for naturalistic settings in experimentation, while still being scientific, the need for risk assessment is increased. For instance in other\footnote{\dots experimentation than that described here.} recent work which we undertook here at the University of Manchester we wished to understand how people used basic mobile phones in a naturalistic setting. This involved the use of SMS text messaging while users were in a mobile setting. Therefore, there was a danger of tripping, falling up or down stairs, walking into obstacles, or into areas of danger. In this case we found it preferable to walk the route noting possible hazards, and solutions, or ways to mitigate those hazards before we applied for ethical approval. However, feedback from the institutional review board pointed out that we had forgotten to include the external dangers; as some of this work was to be conducted in an outside setting there was also a risk of slipping in wet or icy conditions. The review board imposed a restriction to mitigate this risk which stated we could not run the experiments in or after inclement conditions.

Tasks used in a user evaluation depend on the variables tested. If text legibility or reading speed is of concern, text comprehension and word searching tasks are often used. For example, Mustonen et al. examined the affect of walking with regard to  mobile phone text legibility. In their study, participants were asked to read a text passage on a mobile phone and answer questions about the content they had just read~\cite{Mustonen2004}. Participants were also asked to find a given word from a passage of pseudo-text.

Since there is rarely any literature on how to use a small device in a user evaluation, we suggest that our observational results can be used as a default setting. In a user evaluation, users should follow their own typing habit as much as possible. For example, the predictive text function should be turned on, and error correction and use of abbreviations should be allowed. Users should have the right to choose between a physical keyboard and a soft-keyboard. Further, they should not carry additional recording equipment or being closely video recorded or continuously instructed by the experimenter, both of which make the setting unnatural and may affect a participant's performance. The interaction between a user and a experimenter should be kept minimal and only allowed if it is crucial for carrying out the evaluation.

Our suggestion, is that more realistic tasks can be used, such as looking for a telephone number in the contact book, or reading and comprehending a piece of news though the mobile Web. In this way, participants of the study are tested with the tasks that they are familiar with, and therefore the study will be more realistic. On the other hand, if the typing performance of small device users is to be tested, tasks such as composing a text message or an email would be useful. However, one limitation is that since the materials used between participants need to be consistent for the purpose of measuring performance, copy typing is normally used so that all participants type the same text which is given to them before hand. According to our observation, copy typing is not a typical use of small devices. Therefore, a trade-off between the control of variables and the representativeness of real phenomena must be made. Last but not least, pointing performance of small device users is also widely measured. Such tasks normally involve participants clicking on-screen items with a pen and touch-screen. For example, in Brewster et al.'s study, participants were asked to click on-screen buttons of different size with a calculator-style on-screen keyboard~\cite{Brewster2002}. We suggest that dialling a telephone number by clicking the numeric keypad displayed on the device screen, or icon selection for application execution, are a more accurate guide to a user's pointing performance. 

The key aspect of our work is the ability to control the interaction of the participant so that it can be synchronised with the different environmental conditions of the route, coupled with the ability to record the user's interaction with the device. For our experimental designs we decided to use a proxy system such that information was delivered to the mobile device stating what the user should do, and with the facility to carry out those instructions; such as filling in a form with address details, while also recording the interactions with the device. Once the user had completed the action the system would wait for a control signal initiated by the experimenter or observer, following the participant along the route, such that a specific task and action could be associated with a specific stage in the route.  In our case we placed a dummy message saying `loading data to the server' on the user's screen while they were finishing the traversal of one stage of the journey, and before they started another. In this way, the desired action can be sequenced to the environmental and mobile situation. While, at the same time, the various naturalistic traits of mobile usage can be monitored and linked back to the appropriate stage of the mobile traversal; and in addition the log data can be directly linked to the environmental and mobile conditions to which the user is being subjected.

Besides typing habits, small device users also develop attention switch strategies to cope with environmental disruptions. Oulasvirta et al's observation results suggest that small device users normally calibrate their attentions early on, where attention to the environment mainly occurs just when they enter a new environment~\cite{Oulasvirta2005}. When small device users are familiar with the current environment, they will focus more on the device screen and briefly scan the environment over long intervals. Similar observations were also made in our study. We observed that the attention switches of small device users did not spread evenly over the period of observation. They tend to have more attention switches when the environment changed. For example, a small device user had more attention switch when approaching the gate of a shop he wants to enter than when walking from a distance to the shop. In a small device user evaluation, it would be good if multiple environments are used so that the attentional strategies that small device users adopt in real life can apply.

\section{Conclusions}\label{conc}
We have proposed the creation of a set of generic protocols, which can be followed in any mobile experiment mimicking real world mobile usage. Our protocol selection lead us to a number of more general points and experiential aspects. These revolve around user task selection, typing and pointing habits, attention switching strategies, mobility condition selection, and the differentiation between public spaces and in-laboratory routes~\cite{consolvo2007conducting,dahl2010fidelity,po2004heuristic}.

As can be seen from these results, we mainly investigated three parameters that could affect the patterns of use of small devices. These parameters include (i) mobility (move or still, alone or accompanied and silent or talking), (ii) hand usage (one or both hands, thumb or other finger, and keypad or touch screen), and (iii) attention switch (number of attention switches). When we look at our results with these three parameters in mind, our results showed that majority of the observed small device users typed while they were walking, alone, and not talking. This observation is supported by the interview result as majority of the interviewees claimed that they used their small devices while walking and almost half of them reported that they sent text messages while walking. Our results also showed that majority of the observed small device users typed with a physical keypad, using just one hand to manipulate the device, and pressed the keys with thumb. This was also confirmed by the interview results that most of the interviewees typed with one hand, and pressed the keys with thumb.  In terms of the attention switch pattern, observational study results showed that when typing while walking, small device users had 3.27 attention switches in 20 seconds on average. They also had significantly less attention switches when standing still. Interview results indicated that more participants thought they had less attention switches when standing still, which confirmed the observational study results. 

In brief, based on the three parameters discussed above, our two stage study revealed the following patterns of use of small devices:
\begin{itemize}
\item Small device users type on their small devices while they are walking alone and not talking.
\item Small device users type with just one hand, and press the keys with thumb.
\item Small device users use predictive text function, use abbreviations, and correct typing errors in their text messages.
\item Small device users have significantly less attention switches when they are 
typing on the device while not moving than when they do so while walking. 
\item Small device users prefer a physical keyboard to a soft-keyboard.
\end{itemize}

We conclude that researchers following our simple protocols have a much greater chance of uncovering results which more accurately represent small device usage in natural real world settings. We do not claim that our protocols are the finishing point of this type of experimental design~\cite{hagen2005emerging}, but the starting point~\cite{fields2007use}. We do not believe that our work will answer all questions related to controlled experimentation in naturalistic settings but that an informative framework will emerge based on the number of end user studies performed. Therefore, we would encourage the ergonomics community to actively contribute to these protocols such that a progressively more accurate representation of natural experimentation can be created. By making this a community effort we can account for the many and varied mobile situations which might arise, and once created these experimental designs, and the protocols from which they all derived, will enable us to collect and test data far more accurately than is currently the case.
\section{Experimental Data}\label{tech}
\noindent Further details of the study presented here, including the materials, data and its analysis can be found at the Web Ergonomics Lab (WEL) data repository, {\small\url{http://wel-eprints.cs.man.ac.uk/98/}}, and for the confirmatory studies {\small\url{http://wel-eprints.cs.man.ac.uk/118/}}.
\section{References}\label{refs}
\bibliographystyle{spphys}
\bibliography{citeulike-library}

\end{document}